\documentclass[aps,pre,showpacs,floatfix]{revtex4}
\usepackage{epsfig}
\usepackage{amsmath}
\usepackage{amssymb}
\newcommand{\dd}{\text{d}}

\newcommand{\ee}{\text{e}}

\newcommand{\eps}{\varepsilon}

\begin{document}
\title{Casimir effect in the nonequilibrium steady-state of a quantum spin chain}

\author{D. L. Gonz\'alez-Cabrera$^{1,3}$,  Z. R\'acz$^{2}$ and F. van Wijland$^{3}$}

\affiliation{
${}^1$Grupo de F\'{\i}sica T\'eorica de Materia Condensada, Departamento de F\'{\i}sica, Universidad de los Andes, A.A. 4976, Bogot\'a, Colombia\\
${}^2$Institute for Theoretical Physics - HAS, E\"otv\"os University, P\'azm\'any P\'eter s\'et\'any 1/a, 
1117 Budapest, Hungary\\
${}^3$Laboratoire Mati\`ere et Syst\`emes Complexes, CNRS UMR 7057, Universit\'e Paris Diderot -- Paris 7, 10 rue Alice Domon et L\'eonie Duquet, 75205 Paris cedex 13, France\\
}

\date{\today}

\begin{abstract}
We present a fully microscopics-based calculation of the Casimir effect in a nonequilibrium system, namely an energy flux driven quantum XX chain. 
The force between the walls (transverse-field impurities) is calculated 
in a nonequilibrium steady state which is prepared by letting the 
system evolve from an initial state with the two halves of the 
chain prepared at equilibrium at different temperatures. The steady state
emerging in the large-time limit is homogeneous but carries an 
energy flux. The Casimir force in this nonequilibrium state
is calculated analytically in the limit when the transverse fields
are small. We find that the the Casimir force range is reduced 
compared to the equilibrium case, and suggest that the reason 
for this is the reduction of fluctuations in the flux carrying 
steady state. 
\end{abstract}
\maketitle

\section{Introduction}

Historically, the Casimir force \cite{casimir} is the effective interaction that develops between two ideal conductors in the vacuum due to the quantum fluctuations of the electromagetic field. The zero-temperature case was soon generalized by Lifschitz~\cite{lifschitz} who also calculated the effective force induced by thermal fluctuations. Since then, these theoretical predictions have received, as decades elapsed, qualitative and then quantitative confirmations~\cite{Mostepa2001,lamoreaux}. 
For the electromagnetic field, the thermal component of the Casimir force is orders of magnitude weaker than its zero-temperature counterpart. This accounts for the relatively late confirmation of the thermal Casimir force~\cite{lamoreaux,obrechtwildantezzapitaevskiistringaricornell}. The Casimir force has also appeared in other frameworks, such as low dimensional quantum liquids~\cite{fuchsrecatizwerger-1, wachtermedenschonhammer,kolomeiskystraleytimmins}. It was found there, for example, that the Casimir force between magnetic impurities displays Friedel-like oscillations. This large body of the existing literature is mainly devoted to equilibrium situations. 

In this paper, we wish to investigate the interplay of the long range effective interactions produced by an energy flux running through the chain and the fluctuation mediated interactions between two point-like defects located a distance $\ell$ apart. Following the accepted terminology, 
the effective interaction between the defects will be called 
as Casimir force. We have been inspired by a recent series of works by Antezza {\it et al.}~\cite{antezzapitaevskiistringarisvetovoy-1, antezzapitaevskiistringarisvetovoy-2} in which a similar problem was considered, namely the calculation of the Casimir force produced by the electromagnetic 
field in between 
two parallel plates kept at unequal temperatures. This work followed earlier investigations by Polder and Van Hove~\cite{poldervanhove}, recently reviewed by Volokitin and Persson~\cite{volokitinpersson} and Dorofeyev~\cite{dorofeyev}. However, in contrast to \cite{antezzapitaevskiistringarisvetovoy-1, antezzapitaevskiistringarisvetovoy-2}, in our study there will be no phenomenological input in the form of linear response coefficients; it  will be fully microscopicallly based, both for modeling the walls and the bulk of the system. Our 
theoretical laboratory will be an integrable spin chain 
(the XX quantum chain) 
where similar calculations have been carried out before for purely 
equilibrium situations~\cite{fuchsrecatizwerger-1, fuchsrecatizwerger-2, kolomeiskystraleytimmins}. The results we have obtained notably differ from those obtained in a different framework by Antezza {\it et al.}~\cite{antezzapitaevskiistringarisvetovoy-1, antezzapitaevskiistringarisvetovoy-2}. In particular, we find that in constrast to their work, even in the simplest case of two parallel plates, in the presence of an energy flux, the Casimir force does not appear to be the average of the contributions of two equilibrium Casimir forces, corresponding to the two imposed temperatures. We, on the other hand, do not have a simple
decomposition due to the fact that the energy flux is the
quantity which governs the value of the force, and
in our case it clearly emerges that the presence of
the flux decreases the magnitude of the force.

Our results are presented in the following order. Section \ref{TRIjE} briefly reviews the XX chain and describes the properties of nonequilibrium steady-states arising from preparing the system with a step-like temperature profile. In Section \ref{Impurities}, we introduce two magnetic impurities and we show how the steady-state properties are modified in their presence. In Section \ref{Casimir}, we derive the Casimir force exerted by one impurity upon the other and comparison with earlier phenomenological 
calculations are presented. Conclusions and future research directions follow in Section \ref{Conclusion}. Finally, in appendix~\ref{app1} we present a physicist's derivation of the nonequilibrium steady-state properties, previously obtained via more complex $C^*$ algebraic methods.

\section{Transverse XX chain}
\label{TRIjE}
\subsection{Equilibrium}
The transverse XX chain is one of the simplest quantum systems displaying 
long-range correlations and the associated large fluctuations. It is defined by the Hamiltonian
\begin{equation}
{\hat H}=-\sum_i \left( s_i^x s_{i+1}^x + s_i^y s_{i+1}^y + h s_i^z\right)
\label{TRxx}
\end{equation}
where $\vec{s}_i=\frac{1}{2}\vec{\sigma}_i$ and $\sigma_i^\alpha$ ($\alpha=x,y,z$)
denote the three Pauli matrices at sites
$-N, ...,-1, 0,...,N\to\infty $ of a $d=1$ chain, and $h$ is the
transverse field in units of the coupling $J$ and $J=1$ is set in the following.
The standard way~\cite{liebschultzmattis} to study the transverse XX chain is to resort to the Jordan-Wigner transformation which maps the spin chain onto a one-dimensional system of free fermions with energy spectrum $\eps_q=-\cos q-h$:
\begin{equation}
{\hat H}=\sum_q\eps_q c_q^\dagger c_q
\end{equation}
where $c_q$ is the Fourier transform of $c_i=\sigma_i^-\left(\prod_{j\leq i-1}(-\sigma_j^z)\right)$ with the wavenumbers in the range 
$-\pi\le q\le \pi$. As far as large-distance or transport properties are concerned, most of the relevant large scale physics is governed by the modes $q$ in the vicinity of the Fermi level $\pm\kappa$ determined from $\cos \kappa=-h$. We shall thus often resort to the approximation of effective relativistic fermions which consists in linearizing the dispersion relation around $\pm\kappa$ (by setting $q=\pm\kappa+k$) and considering the modes with $q>0$ or $q<0$ as two independent families of relativistic fermions, namely the left and the right movers, with velocity $c=\sin \kappa$. Further details on the validity of such a description can be found e.g. in \cite{tsvelik}. A phenomenological cut-off $\Lambda$ is imposed, when needed, on the new $k$ modes.

\subsection{Steady state with energy flux}
Our goal is to investigate the nonequilibrium states of $\hat H$ which carry
a given energy flux $\langle \hat J_E\rangle \not= 0$. Here the energy flux $\hat{J}_E$ is given by
\begin{equation}
\hat{J_E}=\sum_q\sin q \eps_q c_q^\dagger c_q
\label{eflux}
\end{equation}
One way to achieve a current carrying steady state has recently been discussed by Ogata~\cite{ogata}, who built upon earlier studies by Araki~\cite{araki}, Tasaki~\cite{tasaki}, and Pillet and Aschbacher~\cite{pilletaschbacher}.  Initially, the chain is prepared in the following way: its left side, say from $-\infty$ up to site $j=0$ is in thermal equilibrium at inverse temperature $\beta_1$, while its right hand side from $j=1$ up to $+\infty$ is in equilibrium at inverse temperature $\beta_2$. Both halves are initially disconnected. Then at $t=0$ contact is made through the $(j=0,j=1)$ bond and the system eventually settles into a nonequilibrium steady-state. Due to its infinite thermal conductivity, the temperature profile is flat in the central region whose extent expands at finite velocity. We shall not be concerned here with the dynamics of the formation of the central region and with the
front propagation issues~\cite{raczcourant3, hunyadiraczsasvari, platinikarevski-1, platinikarevski-2}. We focus on the asymptotic 
homogeneous steady-state of the expanding central region of the system, 
where the fermion occupation number in the steady-state is given by
\begin{equation}\label{Fkogata}
\langle c_k^\dagger c_k\rangle=F_k=\frac{\theta(k)}{1+\ee^{\beta_1\eps_k}}+\frac{\theta(-k)}{1+\ee^{\beta_2\eps_k}}
\end{equation}
The above result can be derived mathematically rigorously~\cite{ogata}, a physicist's proof of (\ref{Fkogata}) is provided in appendix~\ref{app1}. It will often prove convenient to introduce $\beta=\frac{\beta_1+\beta_2}{2}$ and $\eps_k'=\theta(k)\frac{\beta_1}{\beta}\eps_k+\theta(-k)\frac{\beta_2}{\beta}\eps_k$, so that $F_k$ appears as the effective Fermi-Dirac occupation number at temperature $\beta$ of free fermions with energy spectrum $\eps_k'$. The discontinuity of $\eps_k'$ is the consequence of the long-range effective interactions in the steady-state, as discussed in \cite{ogata}. Ogata~\cite{ogata} has also discussed the dependence of $\langle\hat{J}_E\rangle$ on $h$, $\beta_1$ and $\beta_2$ in light of the experimental literature~\cite{kudonojikoikenishizakikobayashi}. We note that there exist other ways to produce a nonequilibrium steady-state (see \cite{antalraczsasvari, raczcourant2, raczcourant3}), but we shall not consider these here~\cite{gonzalezcabrera}. 

\section{A spin chain with two impurities}
\label{Impurities}
\subsection{About the force}
In our case, the walls of the Casimir setup will be  
two magnetic impurities at lattice sites $\mp\ell/2$ with strength $\delta h_{1/2}$ which enter the Hamiltonian through
additional terms of the form
\begin{equation}
{\delta\hat H}=-{\delta h}_1 c^\dagger_{-\ell/2}c_{-\ell/2}-{\delta h}_2 c^\dagger_{\ell/2}c_{\ell/2}
\label{dh}
\end{equation}
The task is to determine the effective force between those impurities. Since we are ultimately interested in a nonequilibrium setting, we must circumvent the methods used beforehand in \cite{sundbergjaffe, kolomeiskystraleytimmins, fuchsrecatizwerger-1} that relied on determining a ground-state energy or a free energy. We adopt the following definition of the force on a lattice. Let $\hat{H}_j$ be the energy density at site $j$,
\begin{equation}
\hat{H}_j=-\frac{1}{2}(c_j^\dagger c_{j+1}+c_{j+1}^\dagger c_j)-h c_j^\dagger c_j
\end{equation}
The force $F$ felt by the defect located at site $-\ell/2$ is given by
\begin{equation}\label{force}
F=\frac 12 \langle \hat{H}_{-\ell/2-1}-\hat{H}_{-\ell/2+1}\rangle
\end{equation}
where the average is over the nonequilibrium steady-state. The local energy being quadratic in the fermionic operators, it is clear that in order to calculate $F$, it is sufficient to know the two-point equal time Green's function ${\mathcal G}_{ij}(t,t')=\langle c_i (t) c_j^\dagger(t)\rangle$ as calculated in the next Subsection. 
Note that there is some arbitrariness in defining a force directly on the lattice. In a continuum limit, e.g. when one focuses on wave vectors close to the Fermi level, such differences become irrelevant, and our definition matches the expression of the force one would obtain directly from a continuum theory.

\subsection{Green's functions}
In order to determine the Green's function ${\mathcal G}_{ij}(t,t')=\langle c_i (t) c_j^\dagger(t')\rangle$, we shall need another Green's function $G_{ij}(t,t')$ introduced through the Heisenberg picture,
\begin{equation}
c_i(t)=\sum_n G_{in}(t,0)c_n(0),\;\;c_j^\dagger(t)=\sum_m G^*_{mj}(t,0)c^\dagger_m(0)
\end{equation}
which leads to 
\begin{equation}\label{gcft}{\mathcal G}_{ij}(t,t')=\sum_{m,n}G_{in}(t,0)G^*_{mj}(t',0)\langle c_n (0)c^\dagger_m(0)\rangle
\end{equation}
In the latter formula, we take as the initial state the stationary-state. The Green's function $G$ has to be calculated in the presence of defects, which can be carried out by a variety of methods, one of them being presented in appendix~\ref{app2}. The results can most conveniently be expressed in terms of the Green's function $g_{mn}(\omega)$ in the absence of defects,
\begin{equation}
g_{mn}(\omega)=\int\frac{\dd q}{2\pi}\frac{\ee^{-iq(m-n)}}{-i\omega+\eps_q}
\end{equation}
where $\omega$ is conjugate to $it$. We find that the Fourier transform of $G_{ij}$ reads, in the weak magnetic defect limit,
\begin{equation}\label{GG}\begin{split}
G_{ij}=&g_{ij}+\delta h_1 g_{i-} g_{-j}+\delta h_2 g_{i+} g_{+j}\\&+
\delta h_1^2 g_{i-} g_{-j} g_{--}+\delta h_2^2 g_{i+} g_{+j} g_{++}\\
&+\delta h_1\delta h_2\left( g_{i+} g_{-j} g_{+-}+g_{i-} g_{+j} g_{-+}\right) +{\mathcal O}(\delta h^3)
\end{split}\end{equation}
where $\pm$ is for $\pm\frac{\ell}{2}$. The alternative limit $\delta h_1,\delta h_2\to\infty$ would be corresponding to freezing the degrees of freedom of the defects, which would bring us closer to the original setting of Casimir. In terms of space Fourier transforms, this expression becomes
\begin{equation}\begin{split}
G(q,q',\omega)=&g_q\delta_{q,q'}+\left(\delta h_1\ee^{i(q'-q)\frac{\ell}{2}}+\delta h_2\ee^{-i(q'-q)\frac{\ell}{2}}\right)g_q g_{q'}\\+&\int\frac{\dd k}{2\pi}\left(\delta h_1\ee^{i(q'-k)\frac{\ell}{2}}+\delta h_2\ee^{-i(q'-k)\frac{\ell}{2}}\right)\left(\delta h_1\ee^{i(k-q)\frac{\ell}{2}}+\delta h_2\ee^{-i(k-q)\frac{\ell}{2}}\right)g_qg_{q'}g_k+{\mathcal O}(\delta h^3)
\end{split}\end{equation}
We now use that the steady-state Green's function $\mathcal G$ is given in (\ref{gcft}),
where the initial state $\langle c_n(0) c_m^\dagger(0)\rangle$ is the current-carrying steady-state obtained from a step temperature profile at $t=-\infty$ and  $G$ includes the presence of weak impurities as given in (\ref{GG}). In the infinite time limit, we show in appendix~\ref{app2} that
\begin{equation}\begin{split}
\mathcal G (q,q',t)\simeq &F_Q\delta_{q,q'}-\left[\delta h_1\ee^{i(q-q')\frac{\ell}{2}}+\delta h_2\ee^{-i(q-q')\frac{\ell}{2}}\right]\frac{F_q-F_{q'}}{\eps_q-\eps_q'}\\
&+\int\frac{\dd k}{2\pi}\left(\delta h_1\ee^{i(q'-k)\frac{\ell}{2}}+\delta h_2\ee^{-i(q'-k)\frac{\ell}{2}}\right)\left(\delta h_1\ee^{i(k-q)\frac{\ell}{2}}+\delta h_2\ee^{-i(k-q)\frac{\ell}{2}}\right)\\&\times
\frac{1}{\eps_q-\eps_{q'}}\left[\frac{F_q-F_k}{\eps_q-\eps_k}-\frac{F_{q'}-F_k}{\eps_{q'}-\eps_k}\right]
\end{split}\end{equation}
Since we are interested only in the interactions between the defects, only the cross  $\delta h_1\times \delta h_2$ term contributes to $\mathcal G_\times(q,q')$ when calculating the corresponding force:
\begin{equation}\begin{split}
\mathcal G_\times (q,q')=&
2\delta h_1\delta h_2
\int\frac{\dd k}{2\pi}
\frac{\cos(q+q'-2k)\frac{\ell}{2} }{\eps_q-\eps_{q'}}\left[\frac{F_q-F_k}{\eps_q-\eps_k}-\frac{F_{q'}-F_k}{\eps_{q'}-\eps_k}\right]
\end{split}\end{equation}
It is useful to introduce the function $\gamma(r,\omega)=\int\frac{\dd q}{2\pi}\ee^{iq r}\gamma_q(\omega)$ whose Fourier transform is defined by
\begin{equation}
\gamma_q(\omega)=\frac{1}{-i\omega+\eps'_q}
\end{equation}
With this notation, we have
\begin{equation}\begin{split}
\mathcal G_\times (q,q')=&
2\delta h_1\delta h_2
\int\frac{\dd\omega}{2\pi}\frac{\dd k}{2\pi}
\cos\left[(q+q'-2k)\frac{\ell}{2}\right] \gamma_q(\omega)\gamma_{q'}(\omega)\gamma_k(\omega)
\end{split}\end{equation}
In real space, this becomes
\begin{equation}\begin{split}
\mathcal G_\times (x,y)=&
\delta h_1\delta h_2
\int\frac{\dd\omega}{2\pi}\left[\gamma(\ell/2-x,\omega)\gamma(\ell/2+y,\omega)\gamma(-\ell,\omega)+ \gamma(-\ell/2-x,\omega)\gamma(-\ell/2+y,\omega)\gamma(\ell,\omega)\right]
\end{split}\end{equation}
and this form will be used in the calculation of the force below.

\section{Casimir effect: the effective force between the impurities}
\label{Casimir}
\subsection{Formal expression of the Force}
In this section we derive the expression of the force in the limit where the impurities are a large distance $\ell$ apart. In this limit, many of the expressions encountered above simplify significantly. Most notably, the function $\gamma(r,\omega)$, which enters the final expression of the Green's function, reads
\begin{equation}\label{gcont}
\gamma(r,\omega)=\ee^{i \kappa r} \gamma_1(r,\omega)+\ee^{-i \kappa r} \gamma_2(r,\omega)
\end{equation}
where we used the following definitions
\begin{equation}
\gamma_1(r,\omega)=i\frac{\ee^{-|r\omega|/c_1}}{c_1}\left(\theta(r)\theta(\omega)-\theta(-r)\theta(-\omega)\right)
\end{equation}
and
\begin{equation}
\gamma_2(r,\omega)=i\frac{\ee^{-|r\omega|/c_2}}{c_2}\left(-\theta(r)\theta(-\omega)+\theta(-r)\theta(\omega)\right)
\end{equation}
with $c_1=\frac{\beta_1}{\beta}\sin\kappa$ and $c_2=\frac{\beta_2}{\beta}\sin\kappa$. In equation (\ref{gcont}), the right (1) and left (2) moving fermions account for the $c_1$ and $c_2$ dependent terms, respectively. In terms of $\gamma_1(r,\omega)$ and $\gamma_2(r,\omega)$, the force  can be written as
\begin{equation}\label{force4}\begin{split}
F=&\frac{i\,c_1}{2\,\beta}\left[\sum_{\omega}\left(-\partial_y+\partial_x\right)\ee^{-i\kappa(x-y)}\Gamma_1(x,y,\omega)\right]^{x=y=-\frac{\ell}{2}^+}_{x=y=-\frac{\ell}{2}^-}\\&
-\frac{i\,c_2}{2\,\beta}\left[\sum_{\omega}\left(-\partial_y+\partial_x\right)\ee^{i\kappa(x-y)}\Gamma_2(x,y,\omega)\right]^{x=y=-\frac{\ell}{2}^+}_{x=y=-\frac{\ell}{2}^-}
\end{split}\end{equation}
where the frequency sum is over the $\omega=\frac{2n+1}{\beta}\pi$, and where
\begin{equation}
\Gamma_1(x,y,\omega)=\delta h_1\delta h_2\left(\ee^{-2 i \kappa \ell}\gamma_2(\ell,\omega)\gamma_1(y-\ell/2,\omega)\gamma_1(-x-\ell/2,\omega)+\ee^{2 i \kappa \ell}\gamma_2(-\ell,\omega)\gamma_1(y+\ell/2,\omega)\gamma_1(-x+\ell/2,\omega)\right)
\end{equation}
and
\begin{equation}
\Gamma_2(x,y,\omega)=\delta h_1\delta h_2\left(\ee^{2 i \kappa \ell}\gamma_1(\ell,\omega)\gamma_2(y-\ell/2,\omega)\gamma_2(-x-\ell/2,\omega)+\ee^{-2 i \kappa \ell}\gamma_1(-\ell,\omega)\gamma_2(y+\ell/2,\omega)\gamma_2(-x+\ell/2,\omega)\right)
\end{equation}
Finally, in the limit $\ell\gg 1$, the force between impurities given by Eq.~(\ref{force}) is the discontinuity of the energy density across the defect, and in terms of $\gamma_1(r,\omega)$ and $\gamma_2(r,\omega)$, it can be written as
\begin{equation}\label{force2}\begin{split}
F=&\frac{4\delta h_1\delta h_2\kappa \sin(2\kappa \left|\ell\right|)}{\beta}\sum_{\omega}\gamma_1(\ell,\omega) \gamma_2(-\ell,\omega)\\&
+\frac{2\delta h_1\delta h_2 \cos(2\kappa \ell)}{\beta}\sum_{\omega}\omega \left(\frac{1}{c_1}+\frac{1}{c_2}\right)\gamma_1(\ell,\omega) \gamma_2(-\ell,\omega)
\end{split}\end{equation}
Note that the force can be written in terms of the unperturbed Green's functions of the left and right moving fermions. The expression of the force (\ref{force2}) is the central result of this work. We now specify this result, first to a known situation to make contact with existing results, and second to the physically more interesting case of a current-carrying chain.

\subsection{Equilibrium (recovering existing results)}

In \cite{fuchsrecatizwerger-1} the authors studied the Casimir forces between defects in equilibrium one-dimensional quantum liquids at zero temperature, in order to reproduce their results we have to take the $\beta_1=\beta_2=\beta$ with $\beta\rightarrow\infty$ in equation (\ref{force2}). In this limit the sum over $\omega$ can be changed by an integral $\frac{1}{\beta}\sum_{\omega}\rightarrow\int\frac{d\omega}{2\pi}$, and after some algebra we find
\begin{equation}\label{ftzeroeq}
F=-\delta h_1\delta h_2\left(\frac{\kappa \sin(2\kappa \ell)}{\pi \ell \sin \kappa}+\frac{\cos(2\kappa \ell)}{2\pi \ell^2 \sin \kappa}\right)
\end{equation}
which is indeed the interaction force associated to the interaction potential between impurities given by equation (18) of reference \cite{fuchsrecatizwerger-1}. We conclude that in the case of zero temperature the leading term of the interaction force decays as $1/\ell$ and it oscillates with wavelength $\pi/\kappa$.

\subsection{Out-of-equilibrium, with a heat flux}
In the case $\beta_1\neq\beta_2$  a heat flux  drives the spin chain into a nonequilibrium steady state. The expression for the force reads
\begin{equation}\begin{split}\label{ftnzeroneq}
F=&-\frac{4\delta h_1\delta h_2\kappa \sin(2\kappa \left|\ell\right|)}{\beta c_1 c_2}\left(\frac{1}{\ee^{\frac{\left|\ell\right|\pi p}{\beta}}-\ee^{-\frac{\left|\ell\right|\pi p}{\beta}}}\right)\\&
-\frac{4\pi\delta h_1\delta h_2 p \cos(2\kappa \ell)}{\beta^2 c_1 c_2}\frac{\ee^{\frac{\left|\ell\right|\pi p}{\beta}}}{\left(\ee^{\frac{\left|\ell\right|\pi p}{\beta}}-\ee^{\frac{-\left|\ell\right|\pi p}{\beta}}\right)^2}
\end{split}\end{equation}
where $p=\frac{1}{c_1}+\frac{1}{c_2}$ and we recall that $c_1=\frac{\beta_1 }{\beta}\sin \kappa$ and $c_2=\frac{\beta_2 }{\beta}\sin \kappa$. As one can see force decays  exponentially with a caracteristic length $\xi=\frac{\beta}{\pi p}$. This is in contrast with the zero temperature case where the decay is algebraic. The oscillatory factors  $\sin(2\kappa \left|\ell\right|)$ and $\cos(2\kappa \ell)$ are the same as in the zero temperature limit. As expected, the $\beta_1=\beta_2=\beta$ thermal equilibrium limit is consistent with previously found results~\cite{fuchsrecatizwerger-1}. 



\section{Conclusions}
We have fully determined the Casimir force between two magnetic field defects of weak amplitude  in the nonequilibrium steady-state of the XX spin chain carrying an energy flux. The overall qualitative behavior is similar to that obtained from equilibrium thermal fluctuations: the Casimir force decays exponentially  with  the distance in between the impurities. However, that decay is notably sharper in the presence of an energy flux than without. Thus we conclude that the presence of the energy flux tends to weaken the Casimir force. This can be seen from the correlation length $\xi=\frac{\beta}{\pi p}$ since, as the system approaches equilibrium $\beta_1\to\beta_2$, $\xi_\text{outofeq}-\xi_\text{eq}=-\frac{c(\beta_1-\beta_2)^2}{8 \pi\beta}$, and thus $\xi_\text{out of eq}<\xi_\text{eq}$. This strengthens the general picture~\cite{spohn, eisler} that nonequilibrium fluctuations lead to stiffer systems, which, as the present calculation reveals, do not favor fluctuation mediated interactions.

It is expected that the leading behavior is different in the presence of strong impurities. And so it would be interesting to push our investigations further to probe the similarities and the differences with the calculation of Antezza {\it et al.}~\cite{antezzapitaevskiistringarisvetovoy-2} bearing on the electromagnetic field between two plates thermalized at unequal temperatures. The conceptual issue here is whether different "ensembles" with either a fixed temperature difference or a fixed energy flux should lead to the same physical results in a nonequilibrium setting. Another issue of interest is related to our spin chain having an infinite conductivity. The question of what happens in realistic systems with nonintegrable interactions is an open one which we would like to address in the future.

\label{Conclusion}

\section*{Acknowledgement}
We thank  B. Jancovici for initially drawing the authors' attention to \cite{antezzapitaevskiistringarisvetovoy-1}, Y. Ogata for a useful communication and J.-N. Fuchs for his critical reading of the manuscript. This research has been partially supported by the Hungarian Academy of Sciences (Grant No. OTKA K 68109).

\appendix

\section{Heat baths at different temperatures}\label{app1}
In her work~\cite{ogata}, Ogata resorts to algebraic methods ($C^*$-algebras), to derive the steady-state properties of the spin chain whose two semi-infinite halves are initially prepared in thermalized states at $T_1$ and $T_2$. The goal of the appendix is to rederive her results sticking to the standard methods familiar to physicists.\\

We start from to semi-infinite $XX$ spin chains defined for $n\leq 0$ and $n\geq 1$, which are respectively thermalized at inverse temperatures $\beta_1$ and $\beta_2$. The Hamiltonian for the left hand side reads
\begin{equation}
\hat{H}^{(1)}=\sum_{n\leq -1}\left(-\frac{1}{2}c_n^\dagger c_{n+1}-\frac{1}{2}c_{n+1}^\dagger c_n-h c_n^\dagger c_n\right)=\frac{2}{\pi}\int_0^\pi\dd q\eps_q c_q^\dagger c_q
\end{equation}
where the Fourier transform is given by $c_q=-i\sqrt{\frac{2}{\pi}}\sum_{n\leq 0} \sin [(n-1)q] c_n$ ($0\leq q\leq \pi$) and the energy spectrum is $\eps_q=-\cos q-h$. As for the right hand side we also  have
\begin{equation}
\hat{H}^{(2)}=\sum_{n\geq 0}\left(-\frac{1}{2}c_n^\dagger c_{n+1}-\frac{1}{2}c_{n+1}^\dagger c_n-h c_n^\dagger c_n\right)=\frac{2}{\pi}\int_0^\pi\dd q\eps_q c_q^\dagger c_q
\end{equation}
where the Fourier transform is now given by $c_q=-i\sqrt{\frac{2}{\pi}}\sum_{n\geq 1} \sin [nq] c_n$ ($0\leq q\leq \pi$) and the energy spectrum is of course the same. In the initial state of the left hand side of the chain, we have that
\begin{equation}
\langle c^\dagger_n c_m\rangle=\frac{2}{\pi}\int_0^\pi\dd q\sin [q (n-1)]\sin [q(m-1)] f_q^{(+)},\;n,m\leq 0
\end{equation}
where $f_q^{(1)}=\frac{1}{\ee^{\beta_1\eps_q}+1}$ is the Fermi-Dirac occupation number. Similarly, for the right hand side, we have
\begin{equation}
\langle c^\dagger_n c_m\rangle=\frac{2}{\pi}\int_0^\pi\dd q\sin [q n]\sin [q m] f_q^{(2)},\,n,m\geq 1
\end{equation}
with $f_q^{(2)}=\frac{1}{\ee^{\beta_2\eps_q}+1}$. Ogata's result~\cite{ogata} states that in the steady-state reached in the large time limit, 
\begin{equation}\label{ogatares}
\langle c^\dagger_n c_m\rangle=\int_{-\pi}^{\pi}\frac{\dd k}{2\pi}\ee^{ik(m-n)}F_k
\end{equation} 
where the occupation number $F_k$ is given by
\begin{equation}\label{ogataFk}
F_{k}=\frac{1}{\ee^{\beta_1 \eps_k}+1}\theta(k)+\frac{1}{\ee^{\beta_2 \eps_k}+1}\theta(-k)
\end{equation}
In order to prove the result (\ref{ogataFk}), we start by determining the time dependent Green's function for arbitrary lattice sites. Since $c_k(t)=\ee^{i\eps_k t}c_k(0)$ (the Fourier modes $-\pi\leq k\leq\pi$ refer to the whole translationally invariant chain), we have that 
\begin{equation}
c_n(t)=\int_{-\pi}^\pi\frac{\dd k}{2\pi} \ee^{-ikn}c_k(t)=\int_{-\pi}^\pi\frac{\dd k}{2\pi} \ee^{-ikn+i\eps_k t}\sum_j\ee^{+ikj} c_j(0)
\end{equation}
with a similar relationship for $c^\dagger_m(t)$, hence the Green's function evaluated at equal --but finite-- times reads
\begin{equation}\begin{split}
\langle c_m^\dagger(t) c_n(t)\rangle=&\int_{-\pi}^\pi\frac{\dd k}{2\pi}\frac{\dd k'}{2\pi} \ee^{ik' m-ikn-i\eps_{k'}t+i\eps_k t}\sum_{j,\ell}\ee^{-ik'\ell+ikj} \langle c^\dagger_\ell(0)c_j(0)\rangle
\end{split}\end{equation}
The brackets in the average appearing in the right-hand side denote the sampling with respect to the thermalized initial state. We know that in this initial state
\begin{equation}\begin{split}
\langle c^\dagger_\ell(0)c_j(0)\rangle=\left\{
\begin{array}{ll}
\frac{2}{\pi}\int_0^\pi\dd q \sin [\ell q] \sin [j q] f_q^{(2)}&\text{ if }\ell,j\geq 1\\
\frac{2}{\pi}\int_0^\pi\dd q \sin [(\ell-1)q] \sin [ (j-1) q] f_q^{(1)}&\text{ if }\ell,j\leq 0\\
0&\text{ otherwise}
\end{array}
\right.
\end{split}\end{equation}
It is convenient to rewrite
\begin{equation}
\langle c_m^\dagger(t) c_n(t)\rangle=\langle c_m^\dagger(t) c_n(t)\rangle_{(1)}+\langle c_m^\dagger(t) c_n(t)\rangle_{(2)}
\end{equation}
where
\begin{equation}\begin{split}
\langle c_m^\dagger(t) c_n(t)\rangle_{(1)}=\int_{-\pi}^\pi\frac{\dd k}{2\pi}\frac{\dd k'}{2\pi} \ee^{ik' m-ikn-i\eps_{k'}t+i\eps_k t}\sum_{j,\ell\leq 0}\ee^{-ik'\ell+ikj} \langle c^\dagger_\ell(0)c_j(0)\rangle\\
\langle c_m^\dagger(t) c_n(t)\rangle_{(2)}=\int_{-\pi}^\pi\frac{\dd k}{2\pi}\frac{\dd k'}{2\pi} \ee^{ik' m-ikn-i\eps_{k'}t+i\eps_k t}\sum_{j,\ell\geq 1}\ee^{-ik'\ell+ikj} \langle c^\dagger_\ell(0)c_j(0)\rangle
\end{split}\end{equation}
Let us focus, say, on $\langle c_m^\dagger(t) c_n(t)\rangle_{(2)}$:
\begin{equation}\begin{split}
\langle c_m^\dagger(t) c_n(t)\rangle_{(2)}&=\int_{-\pi}^\pi\frac{\dd k}{2\pi}\frac{\dd k'}{2\pi} \ee^{ik' m-ikn-i\eps_{k'}t+i\eps_k t}\sum_{j,\ell\geq 1}\ee^{-ik'\ell+ikj} \langle c^\dagger_\ell(0)c_j(0)\rangle\\
&=\int_{-\pi}^\pi\frac{\dd k}{2\pi}\frac{\dd k'}{2\pi} \ee^{ik' m-ikn-i\eps_{k'}t+i\eps_k t}\sum_{j,\ell\geq 1}\ee^{-ik'\ell+ikj}\frac{2}{\pi}\int_0^\pi\dd q \sin [\ell q] \sin [j q] f_q^{(2)}\\
\end{split}\end{equation}
Now, we make use of the following identities, valid as $\delta\to 0^+$, in terms of distributions:
\begin{equation}
\sum_{j\geq 1}\sin q j\ee^{\pm i k j}=\lim_{\delta\to 0^+}\frac{1}{2}\frac{\sin q}{\cos k-\cos q\mp i\delta\text{sign}(k)}
\end{equation}
hence we get, the limit $\delta\to 0^+$ being understood, that 
\begin{equation}\begin{split}
\langle c_m^\dagger(t) c_n(t)\rangle_{(2)}&=\int_{-\pi}^\pi\frac{\dd k}{2\pi}\frac{\dd k'}{2\pi} \ee^{ik' m-ikn-i\eps_{k'}t+i\eps_k t}F(k,k')\end{split}\end{equation}
where we have introduced the notation
\begin{equation}\begin{split}
F(k,k')=\frac{2}{\pi}\int_0^\pi\dd q f_q^{(2)}\frac{1}{2}\frac{\sin q}{\cos k-\cos q- i\delta\text{sign}(k)}\frac{1}{2}\frac{\sin q}{\cos k'-\cos q+ i\delta\text{sign}(k')}
\end{split}\end{equation}
We now focus on the $q$ integral that appears in $F(k,k')$~:
\begin{equation}\begin{split}
F(k,k')&=\frac{1}{2}\int_{-\pi}^{\pi}\frac{\dd q}{2\pi}\frac{\sin^2 q f_q^{(2)}}{(\cos k-\cos q- i\delta\text{sign}(k))(\cos k'-\cos q+ i\delta\text{sign}(k'))}\\
&=\frac{1}{\cos k-\cos k'-i\delta(\text{sign}(k)+\text{sign}(k'))}\oint\frac{\dd z}{2\pi i} \sin^2 q f_{q}^{(2)}\\
&\times\left[\frac{1}{1+z^2-2 z(\cos k-i\delta\text{sign}(k))}-\frac{1}{1+z^2-2z(\cos k'+i\delta\text{sign}(k'))}\right]
\end{split}\end{equation}
where we have set $z=\ee^{iq}$ and the $z$-integral runs counter-clockwise around the unit circle. Explicitly carrying out the $z$ integral leads to
\begin{equation}\begin{split}
F(k,k')&=\frac{i}{2}\frac{1}{\cos k-\cos k'-i\delta(\text{sign}(k)+\text{sign}(k'))}\left[f_{k'}^{(2)}\sin k'+f_{k}^{(2)}\sin k\right]
\end{split}\end{equation}
We are thus left with evaluating
\begin{equation}\label{ugly}\begin{split}
\langle c_m^\dagger(t) c_n(t)\rangle_{(2)}&=\int_{-\pi}^\pi\frac{\dd k}{2\pi}\frac{\dd k'}{2\pi} \ee^{ik' m-ikn-i\eps_{k'}t+i\eps_k t}\frac{i}{2}\frac{1}{\cos k-\cos k'-i\delta(\text{sign}(k)+\text{sign}(k'))}\\&
\times \left[f_{k'}^{(2)}\sin k'+f_{k}^{(2)}\sin k\right]
\end{split}\end{equation}
In order to extract the long time behavior of (\ref{ugly}), we change variables from $k$ to  $u= (k-k')t\sin k'$ and we expand in powers of $1/t$ keeping only the leading order at fixed $m$ and $n$:
\begin{equation}\label{o1}\begin{split}
\lim_{t\to\infty}\langle c_m^\dagger(t) c_n(t)\rangle_{(2)}&=\int_{-\pi}^{\pi}\frac{\dd k'}{2\pi} \ee^{ik'(m-n)}f_{k'}^{(2)}i\int_{-\infty}^{+\infty}\frac{\dd u}{2\pi}\frac{\ee^{i u}}{-u-i\text{sign}(k')}\\
&=\int_{-\pi}^{\pi}\frac{\dd k'}{2\pi} \ee^{ik'(m-n)}f_{k'}^{(2)}\theta(-k')
\end{split}\end{equation}
A similar reasoning leads to
\begin{equation}\label{o2}
\lim_{t\to\infty}\langle c_m^\dagger(t) c_n(t)\rangle_{(1)}=\int_{-\pi}^\pi\frac{\dd k}{2\pi}\ee^{ik (m-n) }f_{k}^{(1)}\theta(k)
\end{equation}
Hence, (\ref{o1}) together with (\ref{o2}) are exactly the announced result (\ref{ogatares},\ref{ogataFk}).

\section{Green's function}
\label{app2}

\subsection{Green's functions for the evolution}
In order to determine ${\mathcal G}_{ij}(t,t')$, we introduce the Green's function $G_{ij}(t,t')$ using the Heisenberg picture,
\begin{equation}\label{pourapres}
c_i(t)=\sum_n G_{in}(t,0)c_n(0),\;\;c_j^\dagger(t)=\sum_m G^*_{mj}(t,0)c^\dagger_m(0)
\end{equation}
which leads to ${\mathcal G}_{ij}(t,t')=\sum_{m,n}G_{in}(t,0)G^*_{mj}(t',0)\langle c_n (0)c^\dagger_m(0)\rangle$. In the latter formula, we take as the initial state the stationary-state. The Green's function $G$ in the presence of defects can be determined by a variety of methods. We choose to use a path-integral formulation. Going to Grassmann fields, the Fourier transform of ${\mathcal G}_{ij}(t,t')$ is given by
\begin{equation}
G_{ij}(\omega)=\langle c_i(\omega) \bar{c}_j(\omega)\rangle=\frac{\int{\mathcal D}\bar{c}{\mathcal D}{c}\;c_i(\omega) \bar{c}_j(\omega)\ee^{-S[\bar{c},c]}}{\int{\mathcal D}\bar{c}{\mathcal D}{c}\ee^{-S[\bar{c},c]}}
\end{equation}
where the action $S=S_0+\delta S$ is as follows
\begin{equation}
S_0[\bar{c},c]=\sum_\omega\sum_i \left[(-i\omega -h)\bar{c}_i(\omega) c_i(\omega)-\frac{1}{2} \bar{c}_i c_{i+1}-\frac{1}{2} \bar{c}_ic_{i-1}\right]
\end{equation}
and 
\begin{equation}
\delta S[\bar{c},c]=-\sum_{i,\omega} \delta h_i \bar{c}_i c_i
\end{equation}
and the magnetic field is $\delta h_i=\delta h_+ \delta_{i,-\ell/2}+\delta h_- \delta_{i,\ell/2}$. We introduce $g_{ij}(\omega)$, the Green's function in the absence of defects as given by $S_0$. We start with the definition of the Green's function:
\begin{equation}\begin{split}
\langle c_i(\omega) \bar{c}_j(\omega)\rangle&
=\frac{
\int{\mathcal D}\bar{c}{\mathcal D}{c}\,c_i(\omega) \bar{c}_j(\omega) \ee^{-S[\bar{c},c]}}{\int{\mathcal D}\bar{c}{\mathcal D}{c}  \ee^{-S[\bar{c},c]}}\\
&
=\frac{1}{Z[0,0]}\frac{\delta }{\delta \eta_j}\frac{\delta}{\delta \bar{\eta}_i}Z[\bar{\eta},\eta]\Big|_{\bar{\eta}=\eta=0}
\end{split}\end{equation}
where $\eta$ and $\bar \eta$ are introduced through the partition function:
\begin{equation}\begin{split}
Z[\bar{\eta},\eta]&
=\int{\mathcal D}\bar{c}{\mathcal D}{c} \ee^{-S_0[\bar{c},c]+\sum_{j,\omega}(\delta h_j \bar{c}_j c_j+\bar{\eta}_j c_j+\bar{c}_j \eta_j)}
\end{split}\end{equation}
Then, we use the representation
\begin{equation}
\ee^{\sum_{j,\omega}\delta h_j \bar{c}_{j}c_{j}}=\int\prod_j\mathcal D\bar{\phi}_{j}\mathcal D \phi_{j}\ee^{-\sum_{j,\omega}\left(\bar{\phi}_{j}\phi_{j}+\sqrt{\delta h_j}(\bar{\phi}_{j}c_j+\bar{c}_j \phi_j)\right)}
\end{equation}
giving the partition function as
\begin{equation}\begin{split}
Z[\bar{\eta},\eta]&
=\int\mathcal D\bar{\phi}\mathcal D \phi\ee^{-\sum_{j,\omega}\bar{\phi}_{j}\phi_{j}}
\int{\mathcal D}\bar{c}{\mathcal D}{c} \ee^{-S_0[\bar{c},c]+\sum_{j,\omega}[\bar{c}_j(\eta_j+\sqrt{\delta h_j}\phi_j)+(\bar{\eta}_j+\sqrt{\delta h}_j\bar{\phi}_j)c_j]}\\
&=\int\mathcal D\bar{\phi}\mathcal D \phi\ee^{-\sum_{j,\omega}\bar{\phi}_{j}\phi_{j}}(\det g_{ij})\ee^{\sum_{i,j,\omega}(\bar{\eta}_i+\sqrt{\delta h_i}\bar{\phi}_i)g_{ij}({\eta}_j+\sqrt{\delta h_j}{\phi}_j)}\\
&=\det g_{ij}\ee^{\bar{\eta}_i g_{ij}\eta_j}\int\mathcal D\bar{\phi}\mathcal D \phi\ee^{-\bar{\phi}_i \gamma_{ij}\phi_j+\sum_{j,\omega}(\bar{\xi}_j\phi_j+\bar{\phi}_j\xi_j)}
\end{split}\end{equation}
where we have defined $\gamma_{ij}=\delta_{ij}-\sqrt{\delta h_i\delta h_j}g_{ij}$ and $\xi_m=\sqrt{\delta h_m}\sum_j g_{mj}\eta_j$, $\bar{\xi}_n=\sqrt{\delta h_n}\sum_i g_{in}\bar{\eta}_i$. We perform the remaining path-integral and obtain
 \begin{equation}\label{pf}\begin{split}
Z[\bar{\eta},\eta]&
=\det g_{ij}\det \gamma_{ij}\ee^{\bar{\eta}_i g_{ij}\eta_j+\sum_{n,m,\omega}\bar{\xi}_n[\gamma^{-1}]_{nm}\xi_m}
\end{split}\end{equation}
The Green's function can be now found from the coefficient of $\bar{\eta}_i\eta_j$ in the exponential that appears in(\ref{pf}), so that
\begin{equation}\begin{split}
G_{ij}=&g_{ij}+\sum_{m,n}\sqrt{\delta h_n\delta h_m}g_{in} g_{mj}[\gamma^{-1}]_{nm}\\
=&g_{ij}+\frac{\delta h_+(1-\delta h_- g_{--})g_{i+}g_{+j}+\delta h_-(1-\delta h_+ g_{++})g_{i-}g_{-j}}{(1-\delta h_+ g_{++})(1-\delta h_- g_{--})-{\delta h_+\delta h_-}g_{-+}g_{+-}}\\
&+\frac{{\delta h_+\delta h_-}(g_{i+}g_{-j} g_{+-}+g_{i-}g_{+j}g_{-+})}{(1-\delta h_+ g_{++})(1-\delta h_- g_{--})-{\delta h_+\delta h_-}g_{-+}g_{+-}}
\end{split}\end{equation}
where we used the shorthand notation $\pm$ for $\pm\frac{\ell}{2}$. In the weak magnetic defect limit of interest here we have
\begin{equation}\begin{split}
G_{ij}=&g_{ij}+\delta h_1 g_{i-} g_{-j}+\delta h_2 g_{i+} g_{+j}\\&+
\delta h_1^2 g_{i-} g_{-j} g_{--}+\delta h_2^2 g_{i+} g_{+j} g_{++}\\
&+\delta h_1\delta h_2\left( g_{i+} g_{-j} g_{+-}+g_{i-} g_{+j} g_{-+}\right) +\ldots
\end{split}\end{equation}
It will prove more convenient to resort to the Fourier transforms of these Green functions
\begin{equation}\begin{split}
G(q,q',\omega)=&g_q\delta_{q,q'}+\left(\delta h_1\ee^{i(q'-q)\frac{\ell}{2}}+\delta h_2\ee^{-i(q'-q)\frac{\ell}{2}}\right)g_q g_{q'}\\+&\int\frac{\dd k}{2\pi}\left(\delta h_1\ee^{i(q'-k)\frac{\ell}{2}}+\delta h_2\ee^{-i(q'-k)\frac{\ell}{2}}\right)\left(\delta h_1\ee^{i(k-q)\frac{\ell}{2}}+\delta h_2\ee^{-i(k-q)\frac{\ell}{2}}\right)g_qg_{q'}g_k+\ldots
\end{split}\end{equation}

\subsection{Steady-state Green's frunction}
In order to calculate the steady-state Green's frunction, we start 
with the expression of ${\mathcal G}_{ij}(t,t')$ given right after (\ref{pourapres}) in which the initial state is the current-carrying steady-state obtained from a step temperature profile at $t=-\infty$. After Ogata~\cite{ogata}, if one initially prepares the spin chain with its left hand side at inverse temperature $\beta_-$ and right hand side at inverse temperature 
$\beta_+$, one has that in the steady-state
\begin{equation}
\langle c^\dagger_n(0)c_m(0)\rangle=\int_{-\pi}^\pi\frac{\dd k}{2\pi}\ee^{ik(m-n)} F_k
\end{equation}
with the occupation number $F_k$ given by (\ref{Fkogata}). We now set out to determine $\mathcal G$ explicitly, in the weak impurity strength regimes. In Fourier space we have that
\begin{equation}\begin{split}
\mathcal G (Q,Q',t)=\int\frac{\dd q}{2\pi}G(Q,q,t)G^*(Q',q,t)F_q
\end{split}\end{equation}
In the weak impurity limit, given that
\begin{equation}\begin{split}
G(q,q',t)=&\ee^{i\eps_q t}\delta_{q,q'}-\left[\delta h_1\ee^{i(q-q')\frac{\ell}{2}}+\delta h_2\ee^{-i(q-q')\frac{\ell}{2}}\right]\frac{\ee^{i\eps_q t}-\ee^{i\eps_{q'}t}}{\eps_q-\eps_q'}\\
&+\int\frac{\dd k}{2\pi}\left(\delta h_1\ee^{i(q'-k)\frac{\ell}{2}}+\delta h_2\ee^{-i(q'-k)\frac{\ell}{2}}\right)\left(\delta h_1\ee^{i(k-q)\frac{\ell}{2}}+\delta h_2\ee^{-i(k-q)\frac{\ell}{2}}\right)\\&\times
\frac{1}{\eps_q-\eps_{q'}}\left[\frac{\ee^{i\eps_q t}-\ee^{i\eps_k t}}{\eps_q-\eps_k}-\frac{\ee^{i\eps_{q'}t}-\ee^{i\eps_k t}}{\eps_{q'}-\eps_k}\right]
\end{split}
\end{equation}
and retaining only the leading (time-independent) behavior in the large time limit, we arrive at
\begin{equation}\begin{split}
\mathcal G (Q,Q',t)\simeq &F_Q\delta_{Q,Q'}-\left[\delta h_1\ee^{i(Q-Q')\frac{\ell}{2}}+\delta h_2\ee^{-i(Q-Q')\frac{\ell}{2}}\right]\frac{F_Q-F_{Q'}}{\eps_Q-\eps_Q'}\\
&+\int\frac{\dd k}{2\pi}\left(\delta h_1\ee^{i(Q'-k)\frac{\ell}{2}}+\delta h_2\ee^{-i(Q'-k)\frac{\ell}{2}}\right)\left(\delta h_1\ee^{i(k-Q)\frac{\ell}{2}}+\delta h_2\ee^{-i(k-Q)\frac{\ell}{2}}\right)\\&\times
\frac{1}{\eps_Q-\eps_{Q'}}\left[\frac{F_Q-F_k}{\eps_Q-\eps_k}-\frac{F_{Q'}-F_k}{\eps_{Q'}-\eps_k}\right]
\end{split}\end{equation}
The above formula is at the basis of our calculation of the force.

 

\begin{thebibliography}{0}

\bibitem{casimir}
H. Casimir, Proc. Kon. Nederl. Akad. Wetensch, {\bf B51}, 793 (1948).

\bibitem{lifschitz}
E.M. Lifschitz, Zh. Exsp. Teor. Fiz. {\bf 29}, 94 (1955) [Sov. Phys. JETP {\bf 2}, 73 (1956)]; E.M. Lifschitz and L.P. Pitaevskii, Statistical Physics, Part 2, (Pergamon Press, Oxford, 1980).

\bibitem{Mostepa2001} 
M. Bordag, U. Mohideen, and V.M. Mostepanenko, 
Phys. Rep. {\bf 353}, 1 (2001).

\bibitem{lamoreaux}
S.K. Lamoreaux, Rep. Prog. Phys. {\bf 68}, 201 (2005).

\bibitem{obrechtwildantezzapitaevskiistringaricornell}
J.M. Obrecht, R.J. Wild, M. Antezza, L.P. Pitaevskii, S. Stringari, and E.A. Cornell, Phys. Rev. Lett. {\bf 98}, 063201 (2007).

\bibitem{kolomeiskystraleytimmins}
E.B. Kolomeisky, J.P. Straley, and M. Timmins, Phys. Rev. A {\bf 78}, 022104 (2008).

\bibitem{fuchsrecatizwerger-1}
A. Recati, J.-N. Fuchs, C.S. Peca, and W. Zwerger, Phys. Rev. A. {\bf 72},
023616 (2005).

\bibitem{wachtermedenschonhammer}
P. W\"achter, V. Meden, and K. Sch\"onhammer, Phys. Rev. B {\bf 76}, 045123 (2007).

\bibitem{fuchsrecatizwerger-2} 
J.-N. Fuchs, A. Recati, and W. Zwerger, Phys. Rev. A {\bf 75}, 043615
(2007).

\bibitem{antezzapitaevskiistringarisvetovoy-2}
M. Antezza, L.P. Pitaevskii, S. Stringari, and V.B. Svetovoy, Phys. Rev. A {\bf 77}, 022901 (2008).

\bibitem{poldervanhove}
D. Polder and M. van Hove, Phys. Rev. B {\bf 4}, 3303 (1971).

\bibitem{volokitinpersson}
A.I. Volokitin and B.N.J. Persson, Rev. Mod. Phys. {\bf 79}, 1291 (2007).

\bibitem{dorofeyev}
I.A. Dorofeyev, J. Phys. A {\bf 31}, 4369 (1998).

\bibitem{antezzapitaevskiistringarisvetovoy-1}
M. Antezza, L.P. Pitaevskii, S. Stringari, and V.B. Svetovoy, Phys. Rev. Lett. {\bf 97}, 223203 (2006).

\bibitem{liebschultzmattis} 
E. Lieb, T. Schultz, and D. Mattis, Ann. Phys. (N.Y.) {\bf 16}, 403 (1961).

\bibitem{tsvelik}
A.M. Tsvelik, {\it Quantum field theory in condensed matter physics}, Cambridge University Press, 2003.

\bibitem{ogata}
Y. Ogata, Phys. Rev. E {\bf 66}, 016135 (2002).

\bibitem{araki}
H. Araki, Publ. RIMS, Kyoto Univ. {\bf 20}, 277 (1984).

\bibitem{tasaki}
S. Tasaki, Chaos, Solitons, Fractals {\bf 12}, 2657 (2001).

\bibitem{pilletaschbacher}
C.-A. Pillet and W. Aschbacher, J. Stat. Phys. {\bf 112}, 1153 (2003). 

\bibitem{raczcourant3} 
T. Antal, Z. R\'acz, A. R\'akos, and G.M. Sch\"utz, Phys. Rev. E {\bf 59}, 4912 (1999);
                       Z. R\'acz, J. Stat. Phys. {\bf 101}, 273 (2000).

\bibitem{hunyadiraczsasvari}
V. Hunyadi, Z. R\'acz, and L. Sasv\'ari, Phys. Rev. E {\bf 69}, 066103 (2004).

\bibitem{platinikarevski-2}
T. Platini and D.  Karevski, Journal of Physics: Conference Series {\bf 40}, 93 (2006).

\bibitem{platinikarevski-1}
T. Platini and D. Karevski, Eur. Phy. J. B {\bf 48}, 225 (2005).

\bibitem{antalraczsasvari} 
T. Antal, Z. R\'acz, and L. Sasv\'ari, Phys. Rev. Lett. {\bf 78}, 167 (1997).

\bibitem{raczcourant2} 
T. Antal, Z. R\'acz, A. R\'akos, and G.M. Sch\"utz, Phys. Rev. E {\bf 57}, 5184 (1998).

\bibitem{kudonojikoikenishizakikobayashi}
K. Kudo, T. Noji, Y. Koike, T. Nishizaki, and N. Kobayashi, J. Phys. Soc. Jpn. {\bf 70}, 1448 (2001).


\bibitem{gonzalezcabrera}
D.L. Gonz\'alez Cabrera, PhD Thesis (Universidad de los Andes, Bogot\'a, Colombia, 2009), {\it Out of Equilibrium Systems, Random Matrices and Casimir Effect in Spin Chains}.


\bibitem{sundbergjaffe}
P. Sundberg and R.L. Jaffe, Ann. Phys. (N.Y.) {\bf 309}, 442 (2004).

\bibitem{spohn}
H. Spohn, J. Phys. A {\bf 16}, 4275 (1983).

\bibitem{eisler}
V. Eisler, Z. R\'acz, and F. van Wijland, Phys. Rev. E {\bf 67}, 056129 (2003).


\end{thebibliography}
\end{document}